\documentclass[12pt,a4paper]{article}
\usepackage{a4wide}

\newcommand{\s}[1]{{\rlap/ #1}} 

\begin{document}

\begin{center}

{\Large \bf Bootstrap and the Parameters of Pion-Nucleon Resonances}
\vspace*{0.5cm}

K.~Semenov-Tian-Shanski$^a$,
\underline{A.~Vereshagin}$^{a,b}$ and
V.~Vereshagin$^a$ \\
$^a$St.Petersburg State University, 
$^b$University of Bergen
\vspace*{0.5cm}

{\em Talk given at the X. International Conference On
Hadron Spectroscopy (HADRON'03), \\
August 31 - September 6, 2003,
Aschaffenburg, Germany}

\end{center}

\begin{abstract}
In this talk we demonstrate the results of application of the 
perturbative effective theory formalism developed in papers
\cite{VV} -- \cite{MIN} to the calculation of
$\pi N$ elastic scattering amplitude. Restrictions on the contributing 
resonance parameters are obtained and the low energy coefficients are 
calculated.
\end{abstract}

\section{Introduction}

In
\cite{VV} -- \cite{MIN} it is shown that when working in effective
theory formalism (in the sense of Weinberg), the assumption that the 
perturbation theory (loop expansion for the scattering amplitudes) is 
self consistent, together with the general requirements of covariance, 
unitarity, causality and crossing, leads to certain restrictions for 
the effective Hamiltonian parameters. Moreover, using concrete  
renormalization scheme, it is also possible to obtain constraints (the
{\em bootstrap equations}) for the
{\em physical} parameters of the given amplitude. In other words: one 
can obtain restrictions for the particle spectrum and, thus, perform a 
comparison with the experiment.

We are going to discuss how to obtain those restrictions in case of 
$\pi N$-elastic scattering. As an example, we make the accurate 
estimate of the tensor-to-vector 
$\rho NN$ coupling ratio in complete agreement with the experimental 
data which has never been explained in model-independent way. Besides,
we present the values of the first 48 coefficients in the expansion of 
the tree amplitude around the crossing symmetry point. 

The mathematical background for these calculations and the formalism 
used is reviewed in more details in the talk
\cite{piNAVV}.

\section{$\pi N$ elastic scattering}

The amplitude
$M_{a \alpha}^{b \beta}$ of the reaction
$
\pi_{a} (k) + N_{\alpha} (p, \lambda) \to
\pi_{b} (k') + N_{\beta} (p', \lambda')
$
can be written in the following form:
\[
M_{a \alpha}^{b \beta} = i{(2\pi)}^4 \delta (k+p-k'-p')
\left\{
\delta_{ba}\delta_\alpha^\beta M^+ +
i\varepsilon_{bac}(\sigma_{c})^{\beta\cdot}_{\cdot\alpha}M^{-}
\right\} \;\; .
\]
Here
\[
M^{\pm} = \overline{u}(p',\lambda') \left\{
A^{\pm}+\left(\frac{\s{k}+\s{k'}}{2}\right) B^{\pm}
\right\} u(p,\lambda)\;\; ,
\]
$\s{k} \equiv k_\mu \gamma^\mu$, $a, b = 1,2,3$ and
$\alpha,\beta = 1,2$ stand for the isospin indices,
$\lambda, \lambda'$ --- for polarizations of the initial and final 
nucleons, respectively,
$\overline{u}(p',\lambda')$, $u(p,\lambda)$ --- for Dirac spinors, and
$\sigma_c,\; c=1,2,3$ --- for Pauli matrices. The invariant amplitudes
$A^{\pm}$ and $B^{\pm}$ are the functions of an arbitrary pair of 
scalar kinematical variables
$s \equiv (p+k)^2$, $t \equiv (k-k')^2$, and 
$u \equiv (p-k')^2$.

To construct the 
{\em tree amplitude} one needs to write down the contributions of all 
possible contact vertices and resonance exchange graphs.

We work in the framework of effective theory formalism. This means
that, when constructing the Hamiltonian, we need to take account of
{\em all} the terms consistent with (algebraic) symmetry properties of 
strong interactions; there are no limitations on the number and order 
of field derivatives. Besides, in order to avoid model dependence we 
reserve the possibility to work with infinite number of resonance 
fields and unbounded (though, of course, discrete) mass spectrum.

Altogether this means that the number of items contributing to the
tree level amplitude is actually infinite. This creates a problem: we
have no guiding principle allowing to fix the order of summation. The
way out of this difficulty has been pointed out in
\cite{VV} -- \cite{MIN}. It consists of switching to the
{\em minimal} parametrization for the Hamiltonian and using the method 
of Cauchy forms. The important advantage of this approach is that it 
results in uniformly converging series of pole terms defining the 
amplitude as the polynomially bounded meromorphic function -- no kind 
of singularities but simple poles can appear on this way. To construct 
the Cauchy form for the tree amplitude under consideration, one needs 
to establish the residues (which are the function of coupling 
constants and masses) at the corresponding pole terms (masses) and to 
fix the bounding polynomial degree --- it happens quite sufficient for 
fixing the amplitude up to few unknown functions which, in turn, can 
be found from the
{\em bootstrap equations}

The origin of bootstrap equations is quite natural. Using the 
technique of Cauchy forms, we can get well defined uniformly 
convergent expansions for the invariant amplitudes (we do not write 
them down here due to the lack of space) in three different 
bands on the Mandelstam plane: 
$B_s \{s \sim 0 \}$, $B_t \{t \sim 0 \}$ and 
$B_u \{u \sim 0 \}$. This bands obviously has non-empty intersections
(near the corners of Mandelstam's triangle), and the corresponding 
Cauchy forms are different in each band. Since we need the tree 
amplitude to posses crossing symmetry, each invariant amplitude should 
be a meromorphic function on all the Mandelstam plane. Thus the 
relevant Cauchy forms should coincide in the band intersection 
domains. This results to the set of functional equations (bootstrap 
equations) for the tree level invariant amplitudes, or, the same, to 
infinite set of numerical equations for Hamiltonian parameters%
\footnote{
It is interesting to note that in case of e.g. 
$\pi N$-elastic scattering some of those equations give explicit 
relations between bosonic and fermionic spectrum parameters, thus, 
demonstrating certain 
{\em supersymmetry} features.
}. 

If one uses the renormalized perturbation theory and imposes the 
physical renormalization prescriptions, in which the tree amplitude is 
expressed in terms of 
{\em physical} parameters, then the bootstrap equations becomes the 
restrictions for the physical (measurable) spectrum. In other words,  
the obtained bootstrap equations remains true after renormalization.

It is these equations that can be tested substituting experimental 
data for resonance masses and widths. They also give a possibility to
express one resonance parameter via the other, which, again, can be 
compared with the known data.

\section{Calculation of $G_T/G_V$}

The quantities
$G^T_{NN\rho}$ and 
$G^V_{NN\rho}$ (our minimal parametrization couplings can be related 
to them) were defined and fitted in
\cite{Nagels} as couplings in the following effective Hamiltonian:
\begin{equation}
H^{NN\rho}_{\rm eff} = 
-\overline{N} \left[
G^V_{NN\rho}\gamma_\mu\vec{\rho}^\mu -
G^T_{NN\rho}{\sigma_{\mu\nu}\over 4 m} 
\left(\partial^\mu\vec{\rho}^\nu -\partial^\nu\vec{\rho}^\mu\right)
\right] {1\over 2} \vec{\sigma} N\; ,
\end{equation}
where
$\sigma_a $ are Pauli matrices and
$m$ is the proton mass. 

The existing experimental data
\cite{Nagels} give:
\begin{equation}
{ G^T_{NN\rho} \over G^V_{NN\rho}} \approx 6.1\ ,\;\;
\frac{G_{\pi \pi \rho}G^V_{NN\rho}}{4\pi} \approx 2.4\ , \;\;
G_{\pi \pi \rho} \approx 6.0\ .
\label{gt/gvexp}
\end{equation}

Taking the relevant bootstrap equations (here - 2 of them) we treat 
the above couplings as unknown and express them via other resonance 
parameters%
\footnote{
These particular equations seems to converge fast: among the known 
resonances only
$N(0.94)$, $N(1.44)$, $\Delta (1.23)$ and one meson ---
$\rho (0.77)$, give significant contributions, other possible 
contributions are suppressed by the inverse squares of their mass.  
}, 
the resulting numerical equations being in complete agreement with
(\ref{gt/gvexp}) with
$15\%$ accuracy. 

It should be noted, that the 
$G_T/G_V$ ratio was recently calculated by the authors in the frame of
$KN$-elastic scattering, again, in complete agreement with the 
experiment (to be published).

\section{Low-energy coefficients}

Using the Cauchy forms technique, we have calculated the coefficients 
in the expansion of the 
{\em tree} amplitude around the crossing symmetry point
($t,\, \nu_t \equiv s -u =0$). This coefficients certainly
{\em will} be affected by loop corrections, however, as one can see 
from the 
Table~\ref{table}, 
\begin{table}
\begin{tabular}{l|l|l|l|l|l|l|l|l|} \hline
$\tilde{B}^+$
&$ b^+_{00} $&$ b^+_{01}  $&$ b^+_{02}  $&$ b^+_{03}    $
&$ b^+_{10} $&$ b^+_{11}  $
&$ b^+_{20} $&$ b^+_{21}  $
\\ \hline
{\tiny Experiment}
&$ -3.50    $&$ +0.22     $&$ -0.10     $&$ -0.00036    $
&$ -0.99    $&$ +0.095    $
&$ -0.31    $&$ +0.42     $
\\
&$ \pm 0.10 $&$ \pm 0.10  $&$ \pm 0.05  $&$ \pm0.00004  $
&$ \pm 0.01 $&$ \pm 0.015 $
&$ \pm 0.01 $&$ \pm 0.08  $
\\ \hline
{\tiny Theory}
&$ -4.96    $&$ +0.18     $&$ -0.004    $&$ +0.0001     $
&$ -1.00    $&$ +0.07     $
&$ -0.19    $&$ +0.02     $
\\ \hline\hline
$\tilde{B}^-$
&$ b^-_{00} $&$ b^-_{01}  $&$ b^-_{02}  $&$ b^-_{03}    $
&$ b^-_{10} $&$ b^-_{11}  $
&$ b^-_{20} $&$ b^-_{21}  $
\\ \hline
{\tiny Experiment}
&$ +8.37    $&$ +0.19     $&$ +0.019    $&$ +0.0021     $
&$ +1.08    $&$ -0.063    $
&$ +0.30    $&$ -0.32     $
\\
&$ \pm 0.23 $&$ \pm 0.07  $&$ \pm 0.007 $&$ \pm 0.0002  $
&$ \pm 0.04 $&$ \pm 0.011 $
&$ \pm 0.04 $&$ \pm 0.07  $
\\ \hline
{\tiny Theory}
&$ +8.56    $&$ -0.071    $&$ +0.002    $&$ +0.00003    $
&$ +1.44    $&$ -0.063    $
&$ +0.22    $&$ -0.018    $
\\ \hline\hline
$A^+$
&$ a^+_{00} $&$ a^+_{01}  $&$ a^+_{02}  $&$ a^+_{03}    $
&$ a^+_{10} $&$ a^+_{11}  $
&$ a^+_{20} $&$ a^+_{21}  $
\\ \hline
{\tiny Experiment}
&$ +25.5    $&$ +1.18     $&$ +0.035    $&$ +0.0060     $
&$ +4.60    $&$ -0.051    $
&$ +1.19    $&$ -0.056    $
\\
&$ \pm 0.5  $&$ \pm 0.05  $&$ \pm0.007  $&$ \pm 0.0005  $
&$ \pm 0.12 $&$           $
&$ \pm 0.07 $&$           $
\\ \hline
{\tiny Theory}
&$ +30.2    $&$ +1.1      $&$ +0.04     $&$ +0.007      $
&$ +6.28    $&$ -0.25     $
&$ +1.23    $&$ -0.087    $
\\ \hline\hline
$C^+$
&$ c^+_{00} $&$ c^+_{01}  $&$ c^+_{02}  $&$ c^+_{03}    $
&$ c^+_{10} $&$ c^+_{11}  $
&$ c^+_{20} $&$ c^+_{21}  $
\\ \hline
{\tiny Experiment}
&$ +25.5    $&$ +1.18     $&$ +0.035    $&$ +0.0060     $
&$ +1.12    $&$ +0.15     $
&$ +0.20    $&$ +0.034    $
\\
&$ \pm 0.5  $&$ \pm 0.05  $&$ \pm 0.007 $&$ \pm 0.0005  $
&$ \pm 0.02 $&$ \pm 0.01  $
&$ \pm 0.01 $&$ \pm 0.010 $
\\ \hline
{\tiny Theory}
&$ +30.2    $&$ +1.1      $&$ +0.04     $&$ +0.007      $
&$ +1.3     $&$ -0.10     $
&$ +0.22    $&$ -0.023    $
\\ \hline\hline
$A^-$
&$ a^-_{00} $&$ a^-_{01}  $&$ a^-_{02}  $&$ a^-_{03}    $
&$ a^-_{10} $&$ a^-_{11}  $
&$ a^-_{20} $&$ a^-_{21}  $
\\ \hline
{\tiny Experiment}
&$ -8.87    $&$ -0.34     $&$ +0.1      $&$ -0.0021     $
&$ -1.25    $&$ +0.023    $
&$ -0.338   $&$ +0.305    $
\\
&$          $&$           $&$           $&$             $
&$          $&$           $
&$          $&$           $
\\ \hline
{\tiny Theory}
&$ -9.85    $&$ +0.2      $&$  -0.004   $&$ +0.00007    $
&$ -1.55    $&$ +0.08     $
&$ -0.27    $&$ +0.023    $
\\ \hline\hline
$C^-$
&$ c^-_{00} $&$ c^-_{01}  $&$ c^-_{02}  $&$ c^-_{03}    $
&$ c^-_{10} $&$ c^-_{11}  $
&$ c^-_{20} $&$ c^-_{21}  $
\\ \hline
{\tiny Experiment}
&$ -0.50    $&$ -0.10     $&$ +0.12     $&$ +0.00032    $
&$ -0.17    $&$ -0.039    $
&$ -0.038   $&$ -0.013    $
\\
&$ \pm 0.05 $&$ \pm 0.01  $&$ \pm 0.04  $&$ \pm 0.00003 $
&$ \pm 0.01 $&$ \pm 0.005 $
&$ \pm 0.004$&$ \pm 0.004 $
\\ \hline
{\tiny Theory}
&$ -0.6     $&$ +0.09     $&$ -0.0019   $&$ +0.0001     $
&$ -0.18    $&$ +0.026    $
&$ -0.035   $&$ +0.006    $
\\ \hline\hline
\end{tabular}
\caption{
Low energy coefficients (calculated at the tree level) and their 
experimental values (averaged). In the case of
$A^-$ it is meaningless to calculate errors: the corresponding
quantities are too sensitive to the uncertainties in experimental 
data.
}
\label{table}
\end{table}
the tree level results are very close to the experimental values --- 
this fact gives a hope that our way of constructing the tree amplitude 
\cite{VV}-\cite{MIN} leads to nice convergence of loop expansion, at 
least, in low energy domain. In other words, the tree approximation 
gives nice description of the physical amplitude at low energies%
\footnote{
It should be noted that, in all the cases we checked, the bootstrap 
equations are consistent with the experimental data only if the
tree amplitude asymptotic is taken in accordance with the 
corresponding Regge intersept. In other words, the tree amplitude
{\em shall} have the asymptotic close to the physical one.
}.
 
Introducing the new quantity
\[
C^\pm = A^\pm + {m \nu_t \over 4m^2 - t} \tilde{B}^\pm\; ,
\]
where
$\tilde{B}^{\pm}$ is just
$B^{\pm}$ with the nucleon pole subtracted%
\footnote{
That is what can be compared with the experiment: the nucleon pole 
contribution is dominant in this momentum region but can be excluded 
in experimental data analysis. On the other hand, in our formulas we 
have this contribution explicitly and can simply remove it by hand.
},
we define the low-energy coefficient
$b^{\pm}_{mn}$, $a^{\pm}_{mn}$, and
$c^{\pm}_{mn}$ in the following way:
\[
\tilde{B}^+ (t,\nu_t)
= 
\nu_t \sum_{m,n}^{} b^+_{mn} (\nu_t^2)^m t^n,
\quad
\tilde{B}^- (t,\nu_t)
=
\sum_{m,n}^{} b^-_{mn} (\nu_t^2)^m t^n,
\]
\[
\tilde{A}^+ (t,\nu_t)
=
\sum_{m,n}^{} a^+_{mn} (\nu_t^2)^m t^n,
\quad
\tilde{A}^- (t,\nu_t)
=
\nu_t\sum_{m,n}^{} a^-_{mn} (\nu_t^2)^m t^n,
\]
\[
\tilde{C}^+ (t,\nu_t)
=
\sum_{m,n}^{} c^+_{mn} (\nu_t^2)^m t^n,
\quad
\tilde{C}^- (t,\nu_t)
=
\nu_t\sum_{m,n}^{} c^-_{mn} (\nu_t^2)^m t^n,
\]
where all the expansions are around the point 
$t,\nu_t=0$. Re-expanding corresponding Cauchy forms around this point 
in the above (Taylor) form, using experimental data for couplings and 
masses and neglecting all the contributions of the resonances with
$M\geq 1.9 \;{\rm GeV}$, we get numerical values for the coefficients%
\footnote{
The experimental data can be found in
\cite{Nagels}: please note, that they use somewhat different 
definitions for low-energy coefficient, so one needs to perform 
certain recalculations to compare the results.
}
(see Table~\ref{table}).

Actually, among baryons only
$\Delta(1.23)$ and
$N(1.44)$ give non-negligible contributions as well as
$\sigma$ among mesons, all other known resonances give less then
$10\%$.

\section*{Acknowledgments}

The work was supported in part by INTAS (project 587, 2000), RFBR
(grant 01-02-17152) and by Russian Ministry of Education
(Programme Universities of Russia, project 02.01.001). The work by
A.~Vereshagin was supported by Meltzers H\o yskolefond
(Studentprosjektstipend 2003).




\begin{thebibliography}{9}

\bibitem{VV}
 V.~Vereshagin, 
 {\em Phys. Rev. D} {\bf 55}, 5349 (1997).
\bibitem{AVVV1}
 A.~Vereshagin and V.~Vereshagin,
 {\em $\pi N$ Newsletter} {\bf 15}, 288 (1999).
\bibitem{AVVV}
 A.~Vereshagin and V.~Vereshagin,
 {\em Phys. Rev. D} {\bf 59}, 016002 (2000).
\bibitem{AV}
 A.~Vereshagin,
 {\em $\pi N$ Newsletter} {\bf 16}, 426 (2002).
\bibitem{AVVVKS}
 A.~Vereshagin, V.~Vereshagin, and K.~Semenov-Tian-Shanski, 
 {\em Zap. Nauchn. Sem. POMI} {\bf 291}, Part 17, 78 (2002); English    
 version is to appear in
 {\em J. Math. Sci. (NY)}, (2003).
\bibitem{MIN}
 A.~Vereshagin and V.~Vereshagin, hep-th/0307256, submitted to
 {\em Phys. Rev. D}.
\bibitem{piNAVV}
 K.~Semenov-Tian-Shanski, A.~Vereshagin and V.~Vereshagin,
 ``Bootstrap and the parameters of pion-nucleon resonances'' in
 these Proceedings.
\bibitem{Nagels}
 M.~M.~Nagels et al.,
 {\em Nucl. Phys.} {\bf B109}, 1 (1976);
 O.~Dumbrajs et al., 
 {\em Nucl. Phys.} {\bf B216} 277 (1983).

\end{thebibliography}
\end{document}